\documentclass[11pt]{article}

\usepackage{amsmath}
\usepackage{amssymb}
\usepackage{times}
\usepackage{dsfont}
\usepackage{theorem}

\def\=d{\,{\buildrel\rm def\over =}\,}

\title{Coulomb solutions\\
from\\
improper pseudo-unitary free gauge field operator translations}
\author{Andreas Walter Aste$^{a,b}$\\
$\quad$\\
$^{a}$\emph{Department of Physics, University of Basel, 4056 Basel, Switzerland}\\
$^{b}$\emph{Paul Scherrer Institute, 5232 Villigen PSI, Switzerland}}

\date{August 22, 2014}

\begin{document}
\maketitle

\begin{abstract}
\noindent Fundamental problems of quantum field theory related to the representation
problem of canonical commutation relations are discussed within a gauge field version of a van Hove-type
model.
The Coulomb field generated by a static charge distribution is described as a formal
superposition of time-like pseudo-photons in Fock space with a Krein structure.
In this context, a generalization of operator gauge
transformations is introduced to generate coherent states of abelian gauge fields
interacting with a charged background.\\

\vskip 0.1 cm \noindent {\bf Physics and Astronomy Classification Scheme PACS (2010).}
11.10.-z - Field theory;
11.10.Jj Asymptotic problems and properties;
11.15.-q Gauge field theories;
11.30.-j Symmetry and conservation laws.
\end{abstract}


\section{Introduction}
There is now some interest in occupation numbers of micro-states in classical field configurations
in the context of the entropy of black holes. Some recent works discussed Coulomb fields
as a toy model which connects classical and quantum concepts \cite{Barnich, Mueck}.
However, many of the hitherto presented approaches have
a formal character and neglect mathematical facts and insights which are deeply rooted
in fundamental aspects of quantum field theory.
There is a problem when one wants to count particles in an interacting theory if the
particle notion is based on a Fock space concept and the interaction picture,
as it is expressed by Haag's theorem \cite{Haag}. However, Haag's theorem relies on translation invariance
and does not directly apply to the Coulomb field. Invoking it in the case of \cite{Barnich,Mueck} to draw
any conclusions might therefore be inappropriate.
But it seems that interacting entities are not simply composed of non-interacting entities.
Still, there is an urgent need for the human mind to deconstruct and count the parts of the
surrounding world.\\

\noindent Another aspect of this insight is related to the classification problem of canonical
(anti-)commutation relations and the concept of myriotic fields, since in the quantum
field theoretical case of infinitely many degrees of freedom von Neumann's uniqueness theorem breaks down
\cite{Neumann1, Neumann2, Stone1, Stone2}. It can be shown under natural requirements
that the \emph{formal} canonical commutation relations (CCR) for the position coordinate and conjugate momentum operators
of a physical system with $F$ degrees of freedom
\begin{equation}
[q_l , q_m]=0 \, , \quad [p_l,p_m]=0 \, , \quad [p_l, q_m] =-i \delta_{lm} \, , \quad l,m=1, \ldots F \, ,  \label{ccrn}
\end{equation}
fix the representations of the self-adjoint operators $p_l$, $q_m$ under mild natural requirements
as generators of unitary tranformations
on a Hilbert space up to unitary equivalence, provided $F$ is finite. Already for the case $F=1$ it is straightforward to
show that an algebra fulfilling the commutation relations  eqns. (\ref{ccrn})
cannot be represented by operators defined on a finite-dimensional Hilbert space $\mathcal{H}_f$, since ($\hbar=1$)
\begin{equation}
tr[q,p]=tr(qp) - tr(pq) = 0 \neq i \cdot tr(1)= i \cdot dim(\mathcal{H}_f)  \,  .
\end{equation}
By substituting
\begin{equation}
a_l=(p_l- i q_l)/\sqrt{2} \, , \quad a_l^\dagger = (p_l + i q_l)/\sqrt{2}
\end{equation}
to obtain creation and destruction operators,
one easily derives that the eigenvalues of the occupation operator $N_l=a_l^\dagger a_l$ 
are non-negative integers.
Choosing an occupation number distribution $\{ n \}$ which is an infinite sequence of such integers
in the case $F=\infty$
\begin{equation}
\{ n \} = \{ n_1, n_2, \ldots \} \, ,
\end{equation}
one may divide the set of such sequences into classes such that $\{n \} \sim \{ n' \}$ are in the same
class iff they differ only in a finite number of places.
In the Fock space $\mathcal{F}$, only normalized state vectors
\begin{equation}
N_l \Psi_{\{n^\mathcal{F} \}} = n_l^\mathcal{F}  \Psi_{\{n^\mathcal{F} \}}
\end{equation}
corresponding to an occupation number distribution $\{ n^\mathcal{F} \} $ with
\begin{equation}
\sum_k n_k^\mathcal{F} < \infty
\end{equation}
are allowed to form a complete orthonormal basis in $\mathcal{F}$.
However, an occupation number distribution from a different class $\{ n \} \nsim \{ n^\mathcal{F} \}$
also spans a representation space of the $a_l$, $a_l^\dagger$ and it is evident that representations belonging
to different classes cannot be unitarily equivalent since the creation and destruction operators
change $\{ n \}$ only in one place. An explicit physical example for this problem will be constructed in this
paper. \\

\noindent According to a systematic study concerning the classification of irreducible representations
of canonical (anti-)commutation relations by Garding and Wightman \cite{Garding1,Garding2}, a complete and
practically usable list of representations appears to be unaccessible.
Some interesting comments on the position and momentum operators in wave mechanics
can be found in the appendix.\\

\noindent In flat classical space-time, the proper orthochronous Poincar\'e group $\mathcal{P}_+^\uparrow$
which is a semidirect product of the abelian group of
space-time translations $T_{1,3}$ and the proper orthochronous Lorentz group $\mathcal{L}_+^\uparrow$
\begin{equation}
\mathcal{P}_+^\uparrow = T_{1,3} \rtimes \mathcal{L}_+^\uparrow
\cong T_{1,3} \rtimes SO^+ (1,3) \cong T_{1,3} \rtimes SO(3, \mathds{C})
\end{equation}
is \emph{the} internal symmetry group of the theory.
Relative state phases play an important r\^ole in quantum theory, but
since the global phase of a physical system, represented by a ray in a
Hilbert space, is not observable, the Poincar\'e group ray representations underlying a relativistic
quantum field theory can be realized by necessarily infinite dimensional representations of the covering group
$\bar{\mathcal{P}}^\uparrow_+ \cong T_{1,3} \rtimes SL(2, \mathds{C})$
due to a famous theorem by Wigner \cite{Bargmann, Wigner}.
The actual definition of a particle in non-gravitating flat space-time becomes a non-trivial task when
charged particles coupling to massless (gauge) fields become involved.
Based on the classical analysis of Wigner on the unitary representations of
the Poincar\'e group, a one-particle state describing a particle of mass $m$ alone in the world
is an element of an irreducible representation space of the double cover of the Poincar\'e group in a
physical Hilbert space, i.e. some irreducible representations should occur
in the discrete spectrum of the mass-squared operator $M^2=P_\mu P^\mu$ of a relativistic quantum field theory
describing interacting fields. One should note here that the particles in the present sense like, e.g., a neutron
or an atom, can be viewed as composite objects, and the notion \emph{elementary system} might be more
appropriate. Then, objects like quark and gluons can be viewed as elementary particles, although
they do not appear in the physical spectrum of the Standard Model.
The job of the corresponding elementary fields as carriers of charges is rather to implement the principle of causality
and to allow for a kind of coordinatization of an underlying physical
theory and to finally extract the algebra of observables.
The type and number of the elementary fields appearing in a theory is rather unrelated to the physical spectrum
of empirically observable particles. i.e. elementary systems.
\\

\noindent Furthermore, (idealized) objects like the electron are accompanied by a long range field
which leads an independent life at infinite spatial distance, to give an intuitive picture. It has been shown
in \cite{Buchholz} that a discrete eigenvalue of $M^2$ is absent for states with an electric charge as a direct
consequence of Gauss' law, and one finds that the Lorentz symmetry is not implementable in a sector of
states with nonvanishing electric charge, an issue which also will be an aspect of the forthcoming discussion.
Such problems are related to the fact that the Poincar\'e symmetry
is an overidealization related to global considerations of infinite flat space-time, however, physical measurements have
a local character. The expression \emph{infraparticle} has been coined for charged particles like the electron
accompanied by a dressing field of massless particles \cite{Schroer}.\\

\noindent Still, concentrating on Wigner's analysis of the representations which make sense from a physical
point of view, i.e. singling out tachyonic or negative energy representations and ignoring infraparticle aspects,
the unitary and  irreducible representations of $\bar{\mathcal{P}}^\uparrow_+$ can be classified in the massive case,
loosely speaking, by a real mass parameter $m^2>0$ and a (half-)integer
spin parameter $s$. In the massless case, the unitary irreducible representations of $\bar{\mathcal{P}}^\uparrow_+$
which have played an important r\^ole in quantum field theory so far are those which describe particles with
a given non-negative (half-)integer helicity.\\

\noindent However, one should not forget that there exist so-called infinite spin representations $V_{\Xi,\alpha}$ of
$\bar{\mathcal{P}}^\uparrow_+$  \cite{Yngvason}
which are related to so-called string-localized quantum fields \cite{Mund}.
These representations can be labelled by two parameters $0 < \Xi < \infty$ and $\alpha \in \{ 0, \frac{1}{2} \}$.
The representations describe massless objects with a spin operator along the momentum having the
unbounded spectrum $\{ 0, \pm 1 , \pm 2 , \ldots \}$ for $\alpha=0$ and $\{ \pm \frac{1}{2} , \pm \frac{3}{2} ,
\ldots \}$ for $\alpha=\frac{1}{2}$. The still are ongoing investigations in order to find out whether string-localized
quantum fields will have any direct application in future quantum field theories \cite{Schroer2}. Since the infinite spin representations
can be distinguished by the continuous parameter $\Xi$, they are also called \emph{continuous spin}
representations, a naming which sometimes leads to some confusion about the helicity spectrum which is quantized
but infinite.


\section{The electromagnetic field}
In order to to fix some notational conventions, we shortly mention the well-known fact that Maxwell's equations
in pre-relativistic vector notation
\begin{equation}
div\vec{E}=0 \, , \label{one}
\end{equation}
\begin{equation}
\quad rot \vec{B} - \dot{\vec{E}} = 0 \, , \label{two}
\end{equation}
\begin{equation}
div \vec{B}=0 \, , \label{three}
\end{equation}
\begin{equation}
rot \vec{E} + \dot{\vec{B}} = 0 \, ,  \label{four}
\end{equation}
describing the dynamics of the real classical electromagnetic fields
\begin{equation}
\vec{E}=(E^1,E^2,E^3) \, , \quad
\vec{B}=(B^1,B^2,B^3)
\end{equation}
\emph{in vacuo} can be written by the help of the electromagnetic
field strength tensor $F$ with contravariant components
\begin{equation}
F^{\mu \nu}= -F^{\nu \mu} =
\left(\begin{array}{cccc}
0 & -E^1 & -E^2 & -E^3 \\
E^1 & 0 & -B^3 & B^2 \\
E^2 & B^3 & 0 & -B^1 \\
E^3 & -B^2 & B^1 & 0 
\end{array}\right)\;
\end{equation}
such that eqns. (\ref{one}) and (\ref{two}), which become the inhomogeneous Maxwell equations
in the presence of electric charges, read
\begin{equation}
\partial_\mu F^{\mu \nu}(x) = 0 \, , \quad
\end{equation}
whereas the homogeneous eqns. (\ref{three}) and (\ref{four}) can be written by the help
of the completely antisymmetric Lorentz-invariant Levi-Civita pseudo-tensor $\epsilon$ in four dimensions
with $\epsilon^{0123}=1=-\epsilon_{0123}$
\begin{equation}
\partial_\mu \epsilon^{\mu \nu \rho \sigma} F_{\rho \sigma} (x)= 0 \, .
\end{equation}
Cartesian Minkowski coordinates $x$ have been introduced above where the speed of
light is equal to one such that
$x=(t,\vec{x})=(x^0,x^1,x^2,x^3)=(x_0,-x_1,-x_2,-x_3)$ and $\partial_\mu=\partial/ \partial x^\mu$.\\

\noindent Introducing the gauge vector field or four-vector potential $A$ containing the electrostatic potential $\Phi$
and the magnetic vector potential $\vec{A}$ and skipping space-time arguments for notational simplicity again
\begin{equation}
A^\mu = (\Phi, \vec{A} ) \, , 
\end{equation}
the electric and magnetic fields can be represented via
\begin{equation}
\vec{E} = -grad \, \Phi - \dot{\vec{A}} = -\vec{\nabla} \Phi - \partial_0 \vec{A} \, , \quad \vec{B}= rot \vec{A}
= \vec{\nabla} \times \vec{A} \, , \label{definition}
\end{equation}
or 
\begin{equation}
F^{\mu \nu} = \partial^\mu A^\nu - \partial^\nu A^\mu \, . \label{skew}
\end{equation}
Now, eqns. (\ref{three}) and (\ref{four}) are automatically satisfied by
the definitions in eq. (\ref{definition}), since $rot \, grad \equiv 0$, $div \, rot \equiv 0$
\begin{equation}
rot {E}= -rot \, grad \, \Phi - rot \dot{\vec{A}} = -\dot{\vec{B}} \, , \quad div \vec{B} = div \, rot \vec{A} = 0 \, ,
\end{equation}
and eqns. (\ref{one}) and (\ref{two}) become ($\Box=\partial_\mu \partial^\mu$)
\begin{equation}
\partial_\mu F^{\mu \nu} = \Box A^\nu -\partial^\nu \partial_\mu A^\mu = 0 \, .
\label{problematic}
\end{equation}
Adding the gradient of an arbitrary real analytic scalar field $\chi$  to the gauge field according to
the gauge transformation
\begin{equation}
A^\mu \, \rightarrow \, A^\mu_g = A^\mu + \partial^\mu \chi \label{classical_gauge_trafo}
\end{equation}
leaves $F^{\mu \nu}$ invariant since
\begin{equation}
F^{\mu \nu}_g =\partial^\mu ( A^\nu + \partial^\nu \chi) - \partial^\nu (A^\mu +\partial^\mu \chi) = F^{\mu \nu} \, . 
\end{equation}
One may assume that all fields are analytic and vanish at spatial or temporal infinity rapidly or reasonably fast.
This would exclude global gauge transformations where $0 \neq \chi=const.$ A strong requirement like rapid decrease also
implicitly dismisses infrared problems.
Still, the possibility to perform
a gauge transformation according to eq. (\ref{classical_gauge_trafo}) makes it obvious that eq. (\ref{problematic})
does not fix the dynamics of the gauge field $A^\mu$.
Since for a pure gauge $A^\mu_{pg} = \partial^\mu \chi$
\begin{equation}
\Box \partial^\nu \chi - \partial^\nu \partial_\mu \partial^\mu \chi = 0 \, ,
\end{equation}
the gauge field can be modified in a highly arbitrary manner by the gradient of a scalar function, irrespective
of the initial conditions which define the gauge field on, e.g., a spacelike hyperplane, where the scalar field can
be set to zero. E.g., the zeroth component of eq. (\ref{problematic}) reads
\begin{equation}
\partial_\mu F^{\mu 0 }= \Box A^0 -\partial^0 \partial_\nu A^\nu = -\Delta A^0 - div \dot{A} = div(-grad \Phi - \dot{A}) = div E=0 \, ,
\end{equation}
so there is no equation describing the dynamic evolution of the electrostatic potential $A^0=\Phi$.\\

\noindent The standard way out of this annoying situation in quantum field theory,
where the gauge field is an operator valued distribution, is to modify
eq. (\ref{problematic}) by coupling the four-divergence of
the electromagnetic field strength tensor to an unphysical current term $j_{unph}$,
which in the case of the so-called Feynman gauge is chosen according to
\begin{equation}
\partial_\mu F^{\mu \nu} = \Box A^\nu -\partial^\nu \partial_\mu A^\mu =
-\partial^\nu  \partial_\mu A^\mu = j^\nu_{unph} \, ,
\end{equation}
such that the equations governing the dynamics of the gauge field $A^\mu$ describing a non-interacting
massless spin-1 field from
a more general point of view become
\begin{equation}
\Box A^\mu =0 \, . \label{wave_gauge}
\end{equation}
On the classical level, such a modification can be easily justified by the argument that the four-divergence of the gauge
field $A^\mu$ can be gauged away by a suitable scalar $\chi$ which solves
\begin{equation}
\Box \chi = -\partial_\mu A^\mu \, , \label{inhomog}
\end{equation}
such that for the gauge transformed field $A^\mu_g = A^\mu +\partial^\mu \chi$ one has
\begin{equation}
\partial_\mu A^\mu_g = \partial_\mu (A^\mu + \partial^\mu \chi) = 0 \, .
\end{equation}
Using the retarded propagator $\Delta_0^{ret}$ defined by
\begin{equation}
\displaystyle \Delta_0^{ret}(x)=\int \frac{d^4 k}{(2 \pi)^4} \frac{e^{-ikx}}{k^2+i k^0 0} =
-\frac{1}{2 \pi} \Theta (x^0) \delta(x^2) 
\end{equation}
fulfilling the inhomogeneous wave equation
\begin{equation}
\Box \Delta_0^{ret}(x)=- \delta^{(4)} (x) \, ,
\end{equation}
$\chi$ in eq. (\ref{inhomog}) is given by
\begin{equation}
\chi(x)=\int d^4 x' \,  \Delta_0^{ret}(x-x') \partial_\mu A^\mu (x') + \chi_0(x) 
\end{equation}
with any $\chi_0$ fulfilling $\Box \chi_0(x)=0$.
The formal strategy described above works well even after quantization for QED. However,
when gauge fields couple to themselves, special care is needed.
\\

\noindent In the presence of a conserved four-current $j^\nu$
\begin{equation}
\partial_\mu F^{\mu \nu} = j^\nu \, , \quad \partial_\nu \partial_\mu F^{\mu \nu}= \partial_\nu j^\nu =0
\end{equation}
holds, and invoking the Lorenz condition $\partial_\mu A^\mu=0$ leads to
\begin{equation}
\Box A^\mu = j^\mu \, .
\end{equation}

\noindent The main motivation for the introduction of gauge fields is to maintain explicit locality and manifest covariance
in the quantum field theoretical description of their corresponding interactions. An inversion
of eq. (\ref{skew}) up to a pure gauge is given by
\begin{equation}
A^\mu (x) = \int \limits_0^1 d \lambda \,  \lambda F^{\mu \nu} (\lambda x) x_\nu  \, ,
\end{equation}
but such a term would look rather awkward when substituted in an elegant expression like
the Dirac equation. The Ehrenberg-Siday-Aharonov-Bohm effect \cite{Ehrenberg, Aharonov} also indicates
that the gauge vector field $A$ may play a rather fundamental r$\hat{\mbox{o}}$le in the description of
elementary particle interactions. Many physicists feel that the classical or quantum degrees of freedom of the gauge field $A$
are somehow physical, despite the fact that they are only virtual.
Still, the observable Coulomb field generated by a spherically symmetric charge distribution cannot be composed of real,
asymptotic photons, since such states rather allow for the construction of Glauber states with an electric field
perpendicular to the field momentum.
One also should be cautious to consider a gauge field less physical than the field strength tensor, since
the latter also is no longer gauge invariant in the interacting, non-abelian case. Finally, the
quantum field theoretical Ehrenberg-Siday-Aharonov-Bohm effect is not completely understood as long as no
non-trivial interacting quantum field theory in four space-time dimensions has been constructed at all.\\

\noindent An elegant way to describe the two helicity states of a massless photon
is obtained from combining the electric and magnetic field into a single photon wave function \cite{Silberstein}
\begin{equation}
\Psi=\frac{1}{\sqrt{2}} (\vec{E}+i \vec{B}) \, , \quad i^2=-1  \, . \label{normdef}
\end{equation}
Hence, the Maxwell-Faraday equation
and Amp\`ere's circuital law \emph{in vacuo}
can be cast into the equation of motion
\begin{equation}
\frac{\partial \Psi}{\partial t}  = -i \cdot  \nabla \times \Psi \, . \label{fun}
\end{equation}
This was already recognized in lectures by Riemann in the nineteenth century \cite{Weber}.
Taking the divergence of eq. (\ref{normdef}) 
\begin{equation}
\nabla \cdot \dot{\Psi} = -i \cdot \nabla \cdot (\nabla \times \Psi) = 0
\end{equation}
readily shows that the divergence of the electric and magnetic field is conserved.
Therefore, if the analytic condition
\begin{equation}
div \, \vec{E} = div \,  \vec{B}=0 \label{divcond}
\end{equation}
holds due to the absence of electric or magnetic charges
on a space-like slice of space-time, it holds everywhere.\\

\noindent The field equation (\ref{fun}) and condition (\ref{divcond}) single out the helicity eigenstates of the
photon wave function which are admissible for massless particles
according to Wigner's analysis of the unitary representations of the Poincar\'e group.
E.g., a circularly polarized (right-handed) plane wave moving in {\emph{positive}} $x^3$-direction is given by
\begin{equation}
\Psi_R(x)=N(k^0) \left(\begin{array}{r}
1 \\
i \\
0 \\
\end{array}\right) e^{i k^3 x^3 -i k^0 x^0}
=N(k^0) (\hat{e}_1+ i \hat{e}_2)e^{i k^3 x^3 - i k^0 x^0} \, \, , \quad k^0=k^3 > 0
\end{equation}
where $N(k^0)$ is a normalization factor, whereas the corresponding left-handed plane wave is given by
\begin{equation}
\Psi_L(x)=N(k^0) \left(\begin{array}{r}
1 \\
i \\
0 \\
\end{array}\right) e^{- i k^3 x^3 + i k^0 x^0} \, \, , \quad k^0=k^3 > 0 \, .
\end{equation}
If the right-handed wave moves in negative $x^3$-direction ($k^3<0$), one has
\begin{equation}
\Psi_R(x)=N(k^0) \left(\begin{array}{r}
1 \\
-i \\
0 \\
\end{array}\right) e^{i k^3 x^3 - i k^0 x^0} \, \, , \quad k^0=|k^3| > 0 \, .
\end{equation}
The presence of electric charges and the absence of magnetic charges
breaks the gauge symmetry of eq. (\ref{fun})
\begin{equation}
\Psi \mapsto e^{i \alpha} \Psi \, , \quad \alpha \in \mathds{R} \, .
\end{equation}
\noindent Introducing antisymmetric matrices $ \Sigma_1$, $ \Sigma_2$, and
$ \Sigma_3$ defined by the totally antisymmetric tensor in three dimensions
$\varepsilon_{lmn}=\frac{1}{2}(l-m)(m-n)(n-l)$
\begin{equation}
( \Sigma_l)_{mn} = i \varepsilon_{lmn}
\end{equation}
\begin{equation}
 \Sigma_1=\left(\begin{array}{cccc}
0 & 0 & 0 \\
0 & 0 & i \\
0 & -i & 0 
\end{array}\right)\; , \quad
 \Sigma_2=\left(\begin{array}{cccc}
0 & 0 & -i \\
0 & 0 & 0 \\
i & 0 & 0 
\end{array}\right)\; , \quad
 \Sigma_3=\left(\begin{array}{cccc}
0 & i & 0 \\
-i & 0 & 0 \\
0 & 0 & 0 
\end{array}\right)\; \, ,
\end{equation}
eq. (\ref{fun}) can be written in the form ($\partial_j = \partial / \partial x^j \, ,$ $j=1,2,3$)
\begin{equation}
\frac{\partial \Psi}{\partial t} =  \Sigma_j \partial_j \Psi 
\end{equation}
or, defining matrices $\Gamma^\mu$ by $\Gamma^0=\mathds{1}_3$, where $\mathds{1}_3$ denotes the 
$3 \times 3$  identity matrix, and $\Gamma_j=  \Sigma_j=-\Gamma^j$ for $j=1,2,3,$ eq. (\ref{fun})
finally reads
\begin{equation}
i \Gamma^\mu \partial_\mu \Psi = 0 \, . \label{nice}
\end{equation}
The field components of $\Psi$ covariantly transform under the representation of $SO^+(1,3)$ by the isomorphic
complex orthogonal group $SO(3, \mathds{C})$, preserving condition imposed by eq. (\ref{divcond}).\\

\noindent It has been shown in \cite{Complex} that a mass term for the $\Psi$-field like
\begin{equation}
i \Gamma^\mu \partial_\mu \Psi - m \Psi=0
\end{equation}
is incompatible with the relativistic invariance of the field equation.
As a more general approach one may introduce an (anti-)linear operator S and make the ansatz
\begin{equation}
i \Gamma^\mu \partial_\mu \Psi - m S \Psi=0  \, ,
\end{equation}
which also fails. Already on the classical level, one should be cautious to consider a massless theory as
the limit of a massive theory, which in the case above even does not exist in a naive sense.\\

\noindent What remains in the quantized versions of the classical approaches touched above is the problem
that the use of point-like localized gauge fields is in conflict with the positivity and unitarity of the Hilbert space
and leads to the introduction of Krein structures within a BRS formalism, whereas positivity of the Hilbert space
avoiding unphysical degrees of freedom like in non-covariant Coulomb gauges
necessitates the introduction of a rather awkward non-local formalism.


\section{Lorentz-covariant quantization of the free gauge field}
We quantize the free gauge field as four independent scalar fields in Feynman gauge according to
the canonical commutation relations
\begin{equation}
A^\mu (x)=\frac{1}{(2 \pi)^3} \int \frac{d^3 k}{2 k^0} \bigl[ a^\mu (\vec{k}) e^{-ikx} + a^\mu (\vec{k})^K e^{ikx} \bigr] 
=A^\mu(x)^K \, ,
\end{equation}
where $kx = k_\mu x^\mu =k^0 x^0 - \vec{k} \vec{x} = g_{\mu \nu} k^\mu x^\nu$ and $k^0=k_0=\omega(\vec{k})=| \vec{k} |$
with creation and annihilation operators satisfying
\begin{equation}
[a^\mu(\vec{k}),a^\nu(\vec{k}')^\dagger]=(2 \pi)^3 2 \omega(\vec{k}) \delta^{\mu \nu} \delta^{(3)}(\vec{k}-\vec{k}') \, ,
\end{equation}
\begin{equation}
[a^\mu (\vec{k}),a^\nu (\vec{k}')]=[a^\mu (\vec{k})^\dagger,
a^\nu (\vec{k}')^\dagger]=0
\end{equation}
and all annihilation operators acting on the unique Fock-Hilbert vacuum $| 0 \rangle$ according to
\begin{equation}
a^\mu(\vec{k}) |0\rangle = 0 \, .
\end{equation}
The $K$-conjugation introduced above is necessary due to relativistic covariance and is related to Hermitian conjugation by
\begin{displaymath}
a_0(\vec{k})^K=-a_0(\vec{k})^\dagger \, , \quad a_{1,2,3}(\vec{k})^K=a_{1,2,3}(\vec{k})^\dagger \, ,
\end{displaymath}
\begin{equation}
a_0^\dagger(\vec{k})^K=-a_0(\vec{k}) \, , \quad a_{1,2,3}^\dagger(\vec{k})^K=a_{1,2,3}(\vec{k}) \, ,
\end{equation}
such that the operator valued distributions $A^\mu(x)$ are acting on a Fock-Hilbert space $\mathcal{F}$
with \emph{positive-definite norm}
and since the free field 
\begin{equation}
A^0(x)=\frac{1}{(2 \pi)^3} \int \frac{d^3 k}{2 k^0} \bigl[ a^0 (\vec{k}) e^{-ikx} - a^0 (\vec{k})^\dagger e^{ikx} \bigr]
=-A^0(x)^\dagger
\end{equation}
is anti-Hermitian and due to the commutation relations
\begin{equation}
[a^\mu(\vec{k}),a^\nu(\vec{k}')^K]=-(2 \pi)^3 2 \omega(\vec{k}) g^{\mu \nu} \delta^{(3)}(\vec{k}-\vec{k}') \, ,
\end{equation}
\begin{equation}
[a^\mu (\vec{k}),a^\nu (\vec{k}')]=[a^\mu (\vec{k})^K,
a^\nu (\vec{k}')^K]=0 \, ,
\end{equation}
the gauge field has Lorentz-invariant commutators given by the (positive- and negative-) frequency
Pauli-Jordan distributions $\Delta_0^{(\pm)}$
\begin{equation}
[A^\mu (x),\, A^\nu (y)]=-i g^{\mu\nu} \Delta_0(x-y) \, , \label{photon_commutator}
\end{equation}
with the commutators of the absorption and emission parts alone
\begin{equation}
[A^\mu _-(x),\, A^\nu _+(y)]=-i g^{\mu\nu} \Delta^+_0(x-y) \, , \label{photon_commutator1}
\end{equation}
\begin{equation}
[A^\mu _+(x),\, A^\nu _-(y)]=-i g^{\mu\nu} \Delta^-_0(x-y) \, .\label{photon_commutator2}
\end{equation}
The massless Pauli-Jordan distributions in configuration space are
\begin{equation}
\displaystyle \Delta_0(x)
=-\frac{1}{2 \pi} \mbox{sgn} (x^0) \delta(x^2) \,  ,
\end{equation}
\begin{equation}
\displaystyle  \Delta^\pm_0(x) = \pm \frac{i}{4 \pi^2} \frac{1}{(x_0 \mp i0)^2 -\vec{x}^{\, 2}} \, .
\end{equation}

\noindent Defining the involutive, unitary and Hermitian time-like photon number parity operator $\eta$
defined via the densely defined unbounded photon number operator
\begin{equation}
N_0 = \frac{1}{(2 \pi )^3} \int \frac {d^3 k}{2 k^0} \,  a_0^\dagger (\vec{k}) a_0 (\vec{k})
\end{equation}
by
\begin{equation}
\eta = (-1)^{N_0} = e^{i \pi {N_0}} = e^{-i \pi {N_0}} = \eta^{-1} = \eta^\dagger \, ,
\end{equation}
$\eta$ anticommutes with $a_0(\vec{k})$ and $a_0^\dagger (\vec{k})$, since the creation and annihilation operators
change the time-like particle number by one, and the $K$-conjugation can be defined for an operator $A$ via
\begin{equation}
A^K = \eta A^\dagger \eta \, .
\end{equation}
$\eta$ can be used to define a Krein space $\mathcal{F}_K$ by introducing the indefinite inner product
\cite{Bognar,Krein}
\begin{equation}
(\Phi, \Psi) = \langle \Phi | \eta \Psi \rangle  \, , \quad \Phi, \, \Psi \in \mathcal{F} \, ,
\end{equation}
on $\mathcal{F}$, where $\langle \cdot | \cdot \rangle$ denotes the positive definite scalar
product on the Hilbert space $\mathcal{F}$.


\section{Charge and gauge transformations as field translations}
\noindent Defining the self-adjoint field translation operator $Q$ with four test functions $q_\mu(\vec{k})$, $\mu=0,1,2,3$,
in the Schwartz space of rapidly decreasing functions $ \mathcal{S}(\mathds{R}^3)$ according to
\begin{equation}
Q=\frac{i}{(2 \pi)^3} \int \frac{d^3 k}{2  k^0} \bigl[ q_\nu^* (\vec{k}) a^\nu (\vec{k}) -
q_\nu (\vec{k}) a^\nu (\vec{k})^\dagger \bigr]
\end{equation}
leads to the non-covariant commutation relations
\begin{displaymath}
\bigl[Q, a^\mu (\vec{k}) \bigr] = \frac{i}{(2 \pi)^3} \int \frac{d^3 k'}{2 k'^0}
\bigl[ q_\nu^* (\vec{k}') a^\nu (\vec{k}') -  q_\nu (\vec{k}') a^\nu (\vec{k}')^\dagger , a^\mu(\vec{k}) \bigr]
\end{displaymath}
\begin{equation}
=i \int d^3 k' q_\nu  (\vec{k'})  \delta^{\mu \nu}\delta^{(3)} (\vec{k}-\vec{k}') = i g^{\mu \mu}q^\mu (\vec{k})  
\end{equation}
and
\begin{displaymath}
\bigl[Q, a^\mu (\vec{k})^\dagger \bigr] = \frac{i}{(2 \pi)^3} \int \frac{d^3 k'}{2 k'^0}
\bigl[ q_\nu^* (\vec{k}') a^\nu (\vec{k}') -  q_\nu (\vec{k}') a^\nu (\vec{k}')^\dagger , a^\mu(\vec{k})^\dagger \bigr]
\end{displaymath}
\begin{equation}
= i \int d^3 k' q_\nu^* (\vec{k'}) \delta^{\mu \nu} \delta^{(3)} (\vec{k}-\vec{k}') =  i g^{\mu \mu} q^{\mu} (\vec{k})^*  \, .
\end{equation}

\noindent The $K$-symmetric field translation operator $\tilde{Q}$ defined by
\begin{equation}
\tilde{Q}=\frac{i}{(2 \pi)^3} \int \frac{d^3 k}{2  k^0} \bigl[ q_\nu^* (\vec{k}) a^\nu (\vec{k}) -
q_\nu (\vec{k}) a^\nu (\vec{k})^K \bigr] \label{defK}
\end{equation}
has the commutators
\begin{displaymath}
\bigl[\tilde{Q}, a^\mu (\vec{k}) \bigr] = \frac{i}{(2 \pi)^3} \int \frac{d^3 k'}{2 k^0}
\bigl[ q_\nu^* (\vec{k}') a^\nu (\vec{k}') -  q_\nu (\vec{k}') a^\nu (\vec{k}')^K , a^\mu(\vec{k}) \bigr]
\end{displaymath}
\begin{equation}
= - i\int d^3 k' q_\nu (\vec{k'}) g^{\mu \nu}  \delta^{(3)} (\vec{k}-\vec{k}') =  - i q^\mu (\vec{k})  
\end{equation}
and
\begin{displaymath}
\bigl[\tilde{Q}, a^\mu (\vec{k})^K \bigr] = \frac{i}{(2 \pi)^3} \int \frac{d^3 k'}{2 k^0}
\bigl[ q_\nu^* (\vec{k}') a^\nu (\vec{k}') -  q_\nu (\vec{k}') a^\nu (\vec{k}')^K , a^\mu(\vec{k})^K \bigr]
\end{displaymath}
\begin{equation}
= - i \int d^3 k' q_\nu^* (\vec{k'}) g^{\mu \nu} \delta^{(3)} (\vec{k}-\vec{k}') = - i q^{\mu} (\vec{k})^*  \, .
\end{equation}
Accordingly, one has
\begin{equation}
[ Q,A_0(x)] =\frac{i}{(2 \pi)^3} \int \frac{d^3 k}{2 k^0} [ q_0(\vec{k}) e^{-ikx} - q_0^*(\vec{k}) e^{ikx} ] \, ,
\end{equation}
\begin{equation}
[ Q,A_j (x)] =- \frac{i}{(2 \pi)^3} \int \frac{d^3 k}{2 k^0} [ q_k(\vec{k}) e^{-ikx} + q_k^*(\vec{k}) e^{ikx} ] \, , \quad j=1,2,3 \, ,
\end{equation}
but
\begin{equation}
[ \tilde{Q}, A_0(x)] = - \frac{i}{(2 \pi)^3} \int \frac{d^3 k}{2 k^0} [ q_0(\vec{k}) e^{-ikx} + q_0^*(\vec{k}) e^{ikx} ] \, ,
\end{equation}
\begin{equation}
[ \tilde{Q} ,A_j (x)] =- \frac{i}{(2 \pi)^3} \int \frac{d^3 k}{2 k^0} [ q_k(\vec{k}) e^{-ikx} + q_k^*(\vec{k}) e^{ikx} ] \, , \quad j=1,2,3 \, .
\end{equation}\\

\noindent The operators $Q$ and $\tilde{Q}$ generate unitary and pseudo-unitary transformations $U$ and $\tilde{U}$, respectively
\begin{equation}
U=e^{i Q} \, , \quad \tilde{U}=e^{i \tilde{Q}} \, ,
\end{equation}
with
\begin{equation}
U^\dagger = U^{-1} \, , \quad \tilde{U}^K = \tilde{U}^{-1} \, .
\end{equation}
\noindent The creation and annihilation operators transform according to
\begin{equation}
U a^\mu (\vec{k}) U^{-1} = a^\mu  (\vec{k})+ i [Q, a^\mu (\vec{k})] =
a^\mu (\vec{k}) - g^{\mu \mu} q^\mu (\vec{k}) \, ,
\end{equation}
\begin{equation}
U a^\mu (\vec{k})^\dagger U^{-1} = a^\mu  (\vec{k})^\dagger + i [Q, a^\mu (\vec{k})^\dagger] =
a^\mu (\vec{k})^\dagger - g^{\mu \mu} q^\mu (\vec{k})^* \, ,
\end{equation}
since higher commutator terms vanish in the equations above,
and furthermore
\begin{equation}
\tilde{a}^\mu (\vec{k}) =
\tilde{U} a^\mu (\vec{k}) \tilde{U}^{-1} = a^\mu  (\vec{k}) +
i [\tilde{Q}, a^\mu (\vec{k})] =
a^\mu (\vec{k}) +  q^\mu (\vec{k}) \, ,
\end{equation}
\begin{equation}
\tilde{a}^\mu (\vec{k})^K =
\tilde{U} a^\mu (\vec{k})^K \tilde{U}^{-1} = a^\mu (\vec{k})^K + i [\tilde{Q}, a^\mu (\vec{k})^K] =
a^\mu (\vec{k})^K + q^\mu (\vec{k})^* \, .
\end{equation}

\noindent The vector potential transforms according to
\begin{equation}
A'^\mu (x) = U A^\mu(x) U^{-1} = A^\mu(x) + i [Q,A^\mu(x)]
\end{equation}
and
\begin{equation}
\tilde{A}^\mu (x) = \tilde{U} A^\mu(x) \tilde{U}^{-1} = A^\mu(x) + i [\tilde{Q},A^\mu(x)] \, ,
\end{equation}
i.e. $\tilde{A}^0(x)$ acquires a real expectation value $q^0(x)$ on the Fock vacuum $|0 \rangle$ since
\begin{equation}
\tilde{A}^0 (x) = A^0(x) +  \frac{1}{(2 \pi)^3} \int \frac{d^3 k}{2 k^0} [ q_0(\vec{k}) e^{-ikx} + q_0^*(\vec{k}) e^{ikx} ] 
= A^0(x) + q^0(x) = \tilde{A}^0 (x)^K \, ,
\end{equation}
whereas the unitary transformation $A^0(x) \rightarrow A'^0(x)$ preserves
the skew-adjointness of $A^0$.\\

\noindent One may notice that in the case where $q^\mu(x) = \partial^\mu \chi (x)$ with a smooth scalar
$\chi$ rapidly decreasing in spacelike directions and fulfilling the wave equation $\Box \chi(x) = 0$,
$\tilde{Q}$ becomes a BRST-generator $\tilde{Q}_g$ of free field gauge transformations \cite{Kugo}.
Introducing emission and absorption operators for unphysical photons which are combinations of
time-like and longitudinal states according to
\begin{equation}
b_{1,2}=(a_{\|} \pm a_0)/\sqrt{2}  \,  , \quad
a_{\|}=k_j a^j/|\vec{k}| \, ,
\end{equation}
or
\begin{equation}
b_1=\frac{k_\mu a^\mu}{\sqrt{2} k_0} \, , \quad b_2^\dagger =\frac{ k^\mu  a_\mu^K}{\sqrt{2} k_0} \, ,
\end{equation}
satisfying ordinary commutation relations
\begin{equation}
[b_i (\vec{k}), b^\dagger_j (\vec{k}) ] = (2 \pi)^3 2 k^0 \delta_{ij} \delta^{(3)} (\vec{k} -\vec{k}') \, ,
\end{equation}
one has
\begin{equation}
\partial_\mu A^\mu (x) = - \frac{i}{\sqrt{2} (2 \pi)^3} \int d^3 k \bigl[ b_1 (\vec{k}) e^{-ikx} -
b_2 (\vec{k})^\dagger e^{ikx} \bigr] \, ,
\end{equation}
and
\begin{equation}
b_{1,2}^K = ( a_{||}^\dagger \mp a_0^\dagger)/\sqrt{2} = b_{2,1}^\dagger \, , \quad
\partial^\mu A_\mu^K = \partial^\mu A_\mu \, .
\end{equation}
The free physical sector $\mathcal{F}_{phys} \subset \mathcal{F}$ contains no free unphysical photons
\begin{equation}
| \Phi \rangle \in \mathcal{F}_{phys} \, \,  \Leftrightarrow \, \,
b_1 (\vec{k}) | \Phi \rangle  = b_2 (\vec{k}) | \Phi \rangle = 0 \, \, \forall
\vec{k} \, .
\end{equation}

\noindent A quantum gauge transformation
\begin{equation}
A^\mu_g (x)=A^\mu(x)+\partial^\mu \chi(x) \, , \label{gauge_op}
\end{equation}
with
\begin{equation}
\chi(x)=\int \frac{d^3 k}{(2 \pi)^3 2 k^0} \bigl[ \chi(\vec{k}) e^{-ikx} +
\chi^* (\vec{k}) e^{ikx} \bigr] 
\end{equation}
such that $\chi(x)$ fulfills the wave equation
$\Box \chi(x)=0$
and
\begin{equation}
\partial_\mu \chi(x)=\int \frac{d^3 k}{(2 \pi)^3 2 k^0} \bigl[-i  k_\mu \chi(\vec{k}) e^{-ikx} +
(- i k_\mu \chi (\vec{k}))^* e^{ikx} \bigr] 
\end{equation}
is generated by
\begin{equation}
\tilde{Q}_g = 
- \frac{1}{(2 \pi)^3} \int \frac{d^3 k}{2  k^0} \bigl[ \chi(\vec{k})^*
k_\nu (\vec{k}) a^\nu (\vec{k}) +
\chi (\vec{k}) k_\nu (\vec{k}) a^\nu (\vec{k})^K \bigr] \label{GaugeQ}
\end{equation}
or
\begin{equation}
\tilde{Q}_g = - \frac{1}{\sqrt{2} (2 \pi)^3} \int d^3 k
\bigl[ \chi(\vec{k})^* b_1 (\vec{k}) +
\chi (\vec{k}) b_2^\dagger (\vec{k})  \bigr] \, ,
\end{equation}
\begin{equation}
\tilde{Q}_g = - \frac{1}{\sqrt{2} (2 \pi)^3} \int d^3 k
\bigl[ \chi(\vec{k})^* b_1 (\vec{k}) +
\chi (\vec{k}) b_1^K (\vec{k})  \bigr] \, 
\end{equation}
where $q_\nu (\vec{k})$ has been replaced by $- i k_\nu \chi (\vec{k})$
in eq. (\ref{defK}).
Furthermore, introducing the {\em{gauge current}}
\begin{equation}
j_g^\mu (x) = \! \chi (x)  \! \stackrel {\leftrightarrow}{\partial^\mu} \!
\partial_\nu  A^\nu (x) \, 
\end{equation}
satisfying the continuity equation
\begin{equation}
\partial_\mu j_g^\mu (x) = \partial_\mu (\chi (x) \partial^\mu \partial_\nu A^\nu (x)
- \partial^\mu \chi (x) \partial_\nu A^\nu  (x) )=0 \, ,
\end{equation}
the conserved {\emph{gauge charge}} $\tilde{Q}_g$ can be expressed by \cite{Aste}
\begin{equation}
\tilde{Q}_g=\int \limits_{x^0=const.} d^3x \, j_g^0 (x^0, \vec{x} ) =
\int \limits_{x^0=const.} d^3x \,  \chi (x) \! \stackrel {\leftrightarrow}{\partial^0} \!
\partial_\nu A^\nu (x)    \!\, .
\end{equation}
A generalization of the gauge transformations generated by $\tilde{Q}_g$ to non-abelian
gauge theories including ghost fields has been used in \cite{quantum_gauge}
to derive the classical Lie-structure of gauge theories like QCD from pure quantum principles.
A further generalization to massive QED can be found in \cite{massive_QED},
the Standard Model with a special focus on the elektroweak interaction and the Higgs field
mechanism is discussed in detail in \cite{causalHiggs}.


\section{Static fields}
The field translation operators introduced above modify the free field $A^\mu (x)$ by additional classical
fields $q^\mu (x)$ which are solutions of the wave equation.
 This minor defect if one wants to describe static
fields can be remedied by adding a time-dependence to the classical
$q^\mu$-fields which become $\tilde{q}^\mu(x^0, \vec{k}) = q^\mu (\vec{k}) e^{i k^0 x^0}$.
With the sometimes more suggestive notation $t=x^0$, $\omega=k^0=|\vec{k}|$ and the definitions
\begin{equation}
\tilde{Q}(t)=\frac{i}{(2 \pi)^3} \int \frac{d^3 k}{2  k^0} \bigl[ q_\nu^* (\vec{k}) a^\nu (\vec{k}) e^{-i \omega t}-
q_\nu (\vec{k}) a^\nu (\vec{k})^K e^{+ i \omega t} \bigr] \, ,
\end{equation}
\begin{equation}
\tilde{U}(t) = e^{i \tilde{Q}(t)}
\end{equation}
follows
\begin{equation}
\tilde{A}^0(x) =A^0(x)+  \frac{1}{(2 \pi)^3} \int \frac{d^3 k}{2 k^0} [ q_0(\vec{k}) e^{i \vec{k} \vec{x}} +
q_0^*(\vec{k}) e^{-i \vec{k} \vec{x}} ] 
= A^0(x) + q^0(\vec{x}) \, .
\end{equation}

\noindent The well-known distributional (Fourier transform) identities related to the Coulomb field of a point-like
charge
\begin{equation}
 \int d^3 x \, \frac{e^{\pm i \vec{k}  \vec{x}}}{| \vec{x}|} = \frac{4 \pi}{| \vec{k} |^2} \, , \quad
 \int d^3 x \, e^{\pm i \vec{k} \vec{x}}  \Delta \frac{1}{| \vec{x}|} = - 4 \pi \, , \quad
\Delta \frac{1}{| \vec{x}|} = - 4 \pi \delta^{(3)} (\vec{x}) \, ,
\end{equation}
and
\begin{equation}
V_C (\vec{x}) = \frac{e}{4 \pi} \frac{1}{|\vec{x}|}= \frac{1}{(2 \pi)^3} \int d^3 k \frac{e^{\pm i \vec{k} \vec{x}}}{|\vec{k}|^2}
\end{equation}
can be used to construct a field operator containing a Coulomb field centered at $\vec{x}=0$
as an expectation value ($k^0=|\vec{k}|$)
\begin{displaymath}
A^\mu_c(x)= A^\mu (x) + \delta^\mu_0 \frac{e}{4 \pi} \frac{1}{|\vec{x}|} = A^\mu (x) +  \delta^\mu_0 \frac{1}{(2 \pi)^3}
\int \frac{d^3 k}{k^0} \frac{e^{ i \vec{k} \vec{x}}}{|\vec{k}|}
\end{displaymath}
\begin{equation}
= A^\mu (x) +  \delta^\mu_0 \frac{1}{(2 \pi)^3} \int \frac{d^3 k}{2 k^0}
\Biggl\{ \frac{1}{|\vec{k}|} e^{+i \vec{k} \vec{x}} + \frac{1}{|\vec{k}|} e^{-i \vec{k} \vec{x}} \Biggr\} 
\label{shift}
\end{equation}
fulfilling the inhomogeneous wave equation
\begin{equation}
\Box A^\mu_c (x) = e \delta^{(3)} ( \vec{x}) \, ,
\end{equation}
i.e. one has
\begin{equation}
\tilde{q}^\mu (t, \vec{k}) = (\tilde{q}^{\, 0} (t, \vec{k}), \vec{0} )\, , \quad
\tilde{q}^{\, 0} ( t, \vec{k} ) = \frac{e^{i \omega t}}{| \vec{k} |} \, . \label{timedependent}
\end{equation}\\
The time-dependence of $\tilde{q}^\mu (t, \vec{k})$ could be interpreted as originating
from a kind of binding energy which reduces the energy of non-interacting time-like pseudo-photons
from $\hbar |\vec{k}| c$ to zero when they are bound in a Coulomb field generated by a point-like charge $e$.
In fact, in order to have the correct dynamical time evolution, the
Hamiltonian for non-interacting photons must be
\begin{equation}
H = \frac{1}{(2 \pi)^3} \sum_{\mu=0}^{3}
\int \frac{d^3 k}{2 k^0} \,  \omega(\vec{k}) a_\mu^\dagger (\vec{k}) a_\mu (\vec{k})
=  \frac{1}{2(2 \pi)^3}  \sum_{\mu=0}^{3} \int d^3 k \, a_\mu^\dagger (\vec{k}) a_\mu (\vec{k}) \, ,
\end{equation}
and the improper wave function of a free time-like one-photon state
$| \vec{k}, 0 \rangle = a_0^\dagger (\vec{k}) | 0 \rangle $
is given by
\begin{equation}
\varphi^0_{\vec{k}} (x) = \langle 0 | A^0 (x) | \vec{k}, 0 \rangle =
\langle 0 | A^0 (x)  a_0^\dagger (\vec{k}) | 0 \rangle = 
e^{-i kx} \, ,
\end{equation}
normalized according to
\begin{equation}
\langle \vec{k}, 0 | \vec{k}',0 \rangle = i \int d^3 x \,  {\varphi}^0_{\vec{k}} (x)^*
\stackrel {\leftrightarrow}{\partial_0} {\varphi'}^0_{\vec{k}'} (x) = (2 \pi)^3 2k^0 \delta^{(3)}
(\vec{k}-\vec{k}') \, .
\end{equation}


\section{Particle numbers}
\noindent The field operator $A^\mu_c (x)$ represents a solution of the field equations
for the electromagnetic field interacting with an infinitely heavy point-like charged spinless
particle residing at $\vec{x}=\vec{0}$.
However, the time-like pseudo-photon number operators
\begin{equation}
N_0 = \frac{1}{(2 \pi)^3}
\int \frac{d^3 k}{2 k^0} a_0^\dagger (\vec{k}) a_0 (\vec{k}) \, , 
\quad
\tilde{N}_0(t) = \frac{1}{(2 \pi)^3}
\int \frac{d^3 k}{2 k^0} \tilde{a}_0^\dagger (\vec{k}) \tilde{a}_0 (\vec{k})
\end{equation}
can be written in terms of the untransformed operators as
\begin{displaymath}
(2 \pi)^3 \tilde{N}_0 (t) = \int \frac{d^3 k}{2 k^0} (a_0 (\vec{k})^\dagger + 
\tilde{q}_0 (t, \vec{k})^*) (a_0 (\vec{k})+\tilde{q}_0 (t, \vec{k}))
\end{displaymath}
\begin{equation}
= (2 \pi)^3 N_0 + \int \frac{d^3 k}{2 k^0}
a_0 (\vec{k})^\dagger \tilde{q}_0 (t, \vec{k})
+ \int \frac{d^3 k}{2 k^0}
a_0 (\vec{k}) \tilde{q}_0 (t, \vec{k})^*
+ \int \frac{d^3 k}{2 k^0} 
|\tilde{q}_0 (t, \vec{k})|^2 \, .
\end{equation}
Alternatively, one may write
\begin{equation}
(2 \pi)^3 N_0 = (2 \pi)^3 \tilde{N}_0 (t) - \int \frac{d^3 k}{2 k^0}
\tilde{a}_0 (t,\vec{k})^\dagger \tilde{q}_0 (t, \vec{k})
- \int \frac{d^3 k}{2 k^0} \tilde{a}_0 (t,\vec{k}) \tilde{q}_0 (t, \vec{k})^*
+ \int \frac{d^3 k}{2 k^0} 
|\tilde{q}_0 (t, \vec{k})|^2 \, . \label{number}
\end{equation}

\noindent The displaced vacuum $|\tilde{0}(t) \rangle = \tilde{U}(t) |0 \rangle$
which is time-dependent and permanently modified by the charge,
is \emph{formally} annihilated by the pseudo-unitarily displaced destruction operators $\tilde{a}_0 (\vec{k})$
\begin{equation}
\tilde{a}_0 (t,\vec{k})  |\tilde{0}(t) \rangle =  \tilde{U}(t) a_0 (\vec{k})
\tilde{U}(t)^{-1} \tilde{U}(t) |0 \rangle =  \tilde{U}(t) a_0 (\vec{k}) | 0 \rangle = 0 \, ,
\end{equation}
but it is not Poincar\'e invariant.
It contains infinitely many non-interacting time-like photons, since eq. (\ref{number}) implies
\begin{equation}
\langle \tilde{0} (t) | N_0 | \tilde{0} (t) \rangle = \frac{1}{(2 \pi)^3} \int \frac{d^3 k}{2 k^0} 
|\tilde{q}_0 (t, \vec{k})|^2 \, ,
\end{equation}
which is clearly divergent in the presence of a charge $e \neq 0$.
But one should be cautious about the calculations presented above. In fact, the $\tilde{U}(t)$
are improper pseudo-unitary transformations, since they are not properly defined on the
originally introduced Fock space $\mathcal{F}$.
The pseudo-unitarily inequivalent representations (PUIR) of the canonical quantum field commutation relations
induced by the $\tilde{U}(t)$ relate different spaces at different times.
This also becomes clear if one realizes that the $a_\mu (\vec{k})^\dagger$ are operator valued distributions
\cite{Constantinescu},
such that eq. (\ref{defK}) defines an operator in the sense of a linear operator densely defined on
$\mathcal{F}$ if the $q_\mu$ are Schwartz test functions, i.e. when
the $a_\mu (\vec{k})^\dagger$ are smeared with smooth functions of rapid decrease. Coulomb fields do not
belong to this class of functions.\\

\noindent Screening the Coulomb field \`a la
\begin{equation}
V_C^{scr} (\vec{x}) = \frac{e}{4 \pi | \vec{x}|} e^{-\mu |\vec{x}|} ( 1 - e^{-(m-\mu) | \vec{x} |} ) \,  ,
\quad m \gg \mu > 0 \, , 
\end{equation}
does help, but all divergences reappear in the limit $\mu \rightarrow 0$ or $m \rightarrow \infty$.
Smearing the point-like charge only solves the short-distance (ultraviolet) problems and is related to renormalization
issues in quantum field theory.\\

\noindent It is interesting to note that
\begin{equation}
\langle 0 | \tilde{N}_0 (t) | 0 \rangle = \frac{1}{(2 \pi)^3} \int \frac{d^3k}{2 k^0}
|\tilde{q}_0(t, \vec{k}) |^2 \, ,
\end{equation}
i.e. from the point of view of the theory where the gauge field interacts with an
infinitely heavy charge $e$, the free Fock vacuum $| 0 \rangle$ contains infinitely many
'$\tilde{a}^\dagger$-particles'. Additionally, $\langle \tilde{0} (t) | 0 \rangle =0$ holds
for $t \neq 0$.\\

\noindent Questions concerning the vacuum structure as a ground state in a new physical
sector and a potential non-canonical behaviour of the formal construction above shall not be discussed here.
Still, it should be taken into account that charge screening \emph{is} physical. Considering quantum
electrodynamics restricted to a sector of neutral states with an electromagnetic field decaying faster than
the Coulomb field of a charge distribution with non-zero total charge is fully sufficient to describe the physics
of the photon and charged particles interaction. The scattering process of two electrons is not
really affected by two positrons located very far away, rendering the whole system neutral.
A problem with the description of charged states by local physical operator valued
distributions can be highlighted by the following formal calculation. An operator $C$ carries an elementary
charge $e$, if
\begin{equation}
[\mathcal{Q},C]=e C \, ,
\end{equation}
where $\mathcal{Q}$ denotes the electric charge operator, since if physical states always carry integer multiples of the
elementary charge, they are eigenstates of $\mathcal{Q}$ and $C$ increases the charge of a state $\psi$ with charge $e_0$
by $e$
\begin{equation}
\mathcal{Q} \psi = e_0 \psi \, , \quad \mathcal{Q} C \psi = (e+e_0) C \psi \, ,
\end{equation}
hence
\begin{equation}
\mathcal{Q} C \psi - C \mathcal{Q} \psi = (e+e_0-e_0) C \psi = e C \psi \, .
\end{equation}
From the charge density operator $j_0(x^0,\vec{x})$ one has formally
\begin{equation}
\mathcal{Q} = \lim_{R \rightarrow \infty} \int \limits_{|\vec{x}| < R} j_0(x^0, \vec{x}) \, d^3 x = \lim_{R \rightarrow \infty} \mathcal{Q}_R \, .
\end{equation}
To put it more correctly,  one may consider $j_0(x^0,\vec{x})$ as an operator valued distribution ($x^0=ct, c=1)$
\begin{equation}
\mathcal{Q}_R = \int j_0(x^0,\vec{x}) f_R(\vec{x}) \alpha(t) \, d^3x \, dt 
\end{equation}
with test functions ($\epsilon > 0$)
\begin{equation}
f_R(\vec{x})=f(|\vec{x}|/R) \in \mathcal{D}(\mathds{R}^3) \, , \quad
f(x) = \left\{ \begin{array}{rcl}
1 & : & |x| <  1  \\
0 & : & |x| > 1 + \epsilon \end{array} \right. \quad ,
\end{equation}
\begin{equation}
\alpha(t) \in \mathcal{D}(\mathds{R}) \, , \quad \int \limits_{-\infty}^{\infty} \alpha(t) dt =1 \, .
\end{equation}
Insisting on Gauss' law for local physical operator-valued distributions describing electric currents and electromagnetic fields
\begin{equation}
j_\mu (x) = \partial^\nu F_{\nu \mu}(x) 
\end{equation}
implies by partial integration
\begin{displaymath}
\lim_{R \rightarrow \infty} [\mathcal{Q}_R,C]= \lim_{R \rightarrow \infty} \int d^3 x \, dt \, f_R(\vec{x}) \alpha(t)
[ \partial^j F_{j0} (t,\vec{x}), C ] 
\end{displaymath}
\begin{equation}
= \lim_{R \rightarrow \infty} \int d^3 x \, dt \, \partial_j f_R(\vec{x}) \alpha(t) [ F_{j0} (t,\vec{x}), C ] \, .
\label{commcharge}
\end{equation}
But $\vec{\nabla} f_R(\vec{x}) \neq 0$ only holds for $R < |\vec{x}| < R(1+\epsilon)$.
For local field operators
\begin{equation}
C=\int d^4x \, C(x) g(x) \, , \quad g \in \mathcal{D}(\mathds{R}^4)
\end{equation}
one has a for sufficiently large $R$ a space-like separation of the supports
$\mbox{supp}(\vec{\nabla} f_R (\vec{x}) \alpha(t)   )$ and $\mbox{supp}(g)$.
Due to causality, one has from the vanishing commutators ($\vec{x} \in \mbox{supp}(\vec{\nabla}f_R)$) and eq. (\ref{commcharge})
\begin{equation}
 [ F_{j0} (t,\vec{x}), C ] \overset{R \rightarrow \infty}{\rightarrow} 0 \, , \quad \mbox{i.e.} \quad [\mathcal{Q}, C ] =\lim_{R \rightarrow \infty}
[\mathcal{Q}_R, C]=0 \, ,
\end{equation}
hence $C$ is uncharged. The argument above also works for test functions of rapid decrease in
Schwartz spaces $\mathcal{S}(\mathds{R}^{n})$. An electron alone in the world cannot be created, and an 
accompanying infinitely extended Coulomb field does not exist. Many problems in the quantum field theory stem from the overidealization
that by translation invariance extended systems are considered over infinite space and time regions. However, counting
(unphysical) photons in a restricted space region might make sense.

\section{Conclusions}
\noindent
Already in 1952, van Hove investigated a model where a neutral scalar field interacts with
a source term describing an infinitely heavy, recoilless or static point-like nucleon \cite{vanHove}.
There he showed that the Hilbert space of the free scalar field is 'orthogonal' to the Hilbert
space of states of the field interacting with the point source. His finding finally lead to what is called
today Haag's theorem. This theorem has been formulated in different versions, but basically
it states that there is no proper unitary operator that connects the Fock representation
of the CCR of a non-interacting quantum field theory with the Hilbert space of a corresponding
theory which includes a non-trivial interaction. Furthermore, the interacting Hamiltonian
is not defined on the Hilbert space on which the non-interacting Hamiltonian is defined.
The present paper generalizes van Hove's model by including gauge fields. One should remark that
gauge transformations generated by an operator like $\tilde{Q}_g$ in eq. (\ref{GaugeQ})
can be implemented on one Fock-Hilbert space since smearing free field operators with test functions
defined on a three-dimensional spacelike plane already gives some well-defined operators,
although the fields a operator-valued distributions on four-dimensional Minkowski space.\\

\noindent A formal way out of the lost cause of unitary inequivalent representations (UIR)
is possibly provided by causal perturbation theory
introduced in a classic paper by Epstein and Glaser \cite{EpsteinGlaser}.
In the traditional approach to quantum field theory, one starts
from classical fields and a Lagrangean which includes distinguished interaction terms.
The formal free field part of the theory gets quantized and perturbative $S$-matrix elements or
Greens functions are constructed with the help of the Feynman rules based on a Fock space
description. E.g., a typical model theory often used in theoretical considerations
is the (massless) $\Phi^3$-theory, where the interaction
Hamiltonian density is given by the normally ordered third order monomial
of a free uncharged (massless) scalar field and a coupling
constant $\lambda$
\begin{equation}
{\cal{H}}_{int}(x)=\frac{\lambda}{3!} : \Phi(x)^3:.
\end{equation}
The perturbative $S$-Matrix is then constructed according to the expansion
\begin{equation}
S={\bf 1}+\sum \limits_{n=1}^{\infty} \frac{(-i)^n}{n !} \int dx^4_1 ... dx^4_n
T\{{\cal{H}}_{int}(x_1) {\cal{H}}_{int}(x_2) \cdot ... \cdot {\cal{H}}_{int}(x_n)\}, \label{stoer1}
\end{equation}
where $T$ is the time-ordering operator.
It must be pointed out that the perturbation series eq. (\ref{stoer1}) is formal
and it is difficult to make any statement about the convergence of this series,
but it is erroneously hoped that $S$ reproduces the full theory.\\

\noindent On the perturbative level, two problems arise in the expansion given above.
First, the time-ordered products
\begin{equation}
T_n(x_1,x_2, ...,x_n)=(-i)^n T\{{\cal{H}}_{int}(x_1) {\cal{H}}_{int}(x_2)
\cdot ... \cdot {\cal{H}}_{int}(x_n)\} \label{toprod}
\end{equation}
are usually plagued by ultraviolet divergences. However, these divergences can be removed by
regularization to all orders if the theory is renormalizable,
such that the operator-valued distributions $T_n$ can be viewed as
well-defined, already regularized expressions.
Second, infrared divergences are also present in eq. (\ref{stoer1}).
This is not astonishing, since the $T_n$'s are operator-valued distributions, and
therefore must be smeared out by test functions in $\mathcal{S}(\mathds{R}^{4n})$.
One may therefore introduce
a test function $g(x) \in \mathcal{S}(\mathds{R}^{4})$ which plays the role of an
'adiabatic switching' and provides a cutoff in the long-range part of the interaction, which
can be considered as a natural infrared regulator \cite{EpsteinGlaser,PPNP}.
Then, according to Epstein and Glaser, the infrared regularized $S$-matrix is given by
\begin{equation}
S(g)={\bf 1}+\sum \limits_{n=1}^{\infty} \frac{1}{n !} \int dx^4_1 ... dx^4_n
T_n(x_1,...,x_n) g(x_1) \cdot ... \cdot g(x_n), \label{smoothedS}
\end{equation}
and an appropriate adiabatic limit $g \rightarrow 1$ must be performed at the end
of actual calculations in the right quantities (like cross sections)
 where this limit exists. 
This is not one of the standard strategies usually
found in the literature, however, it is the most natural one
in view of the mathematical framework used in perturbative quantum field
theory.
From a non-perturbative point of view, one may hope that
taking matrix elements in the right quantities allows to reconstruct the full interacting
Hilbert space.
\\

\noindent D\"utsch, Krahe and Scharf performed perturbative calculations for
electron scattering off an electrostatic potential in the framework of causal perturbation theory
\cite{DKS}.
It was found that in the adiabatic limit
$g \rightarrow 1$ the electron scattering cross section is unique only if in the soft bremsstrahlung contributions
from all four photon polarizations are included. Summing over two physical polarizations only,
non-covariant terms survive in the physical observables. The adiabatic switching
in the causal approach has the unphysical consequence that electrons lose their charge
in a distant space-time region. This switching is moved from our local reality to infinity
in the limit $g \rightarrow 1$. As long as $g \neq 1$, the decoupled photon field is no longer
transversal, but also consists of scalar and longitudinal photons. In reality, these photons
are confined to the charged particles and make them charged. Hence, there is some
kind of confinement problem in QED.
One must conclude that
counting unphysical objects is a delicate task.

\section*{Appendix: Position and momentum operators}
In the case $F=1$, the position and conjugate momentum operators
$q$ and $p$ cannot be both bounded linear operators defined everywhere on a
separable Hilbert space $\mathcal{H}$ \cite{Wintner, Wielandt},
since in this case operator norms defined by the Hilbert space norm $|| \cdot ||$
induced by the scalar product on $\mathcal{H}$
\begin{equation}
||q||=\sup_{\Psi \in \mathcal{H}}
\frac{||  q \Psi ||}{||\Psi||} \, , \quad
 ||p||=\sup_{\Psi \in \mathcal{H}} \frac{|| p \Psi ||}{||\Psi||}
\end{equation}
would exist and by induction
\begin{equation}
[q,p^2] =p[q,p]+[q,p]p=2ip \, ,
\end{equation}
\begin{equation}
[q,p^3]=p[q,p^2]+[q,p]p^2 = 3ip^2 \, , \, \ldots
\end{equation}
\begin{equation}
[q,p^n]=inp^{n-1} \label{contraindication}
\end{equation}
would hold for $n \ge 1$.
Since the operator norm is submultiplicative (e.g., $||qp|| \leq ||q|| \, ||p||$) and due to the Cauchy-Schwarz
inequality, eq. (\ref{contraindication}) would imply
\begin{equation}
n ||p^{n-1}||=||[q,p^n]|| \le 2 ||q|| \, ||p^n|| \le 2 ||q|| \, ||p|| \, ||p^{n-1}|| \le C ||p^{n-1}|| 
\end{equation}
for some constant $C$.
For $n > C$ follows $p^{n-1}=0$, and finally one is successively lead to a contradiction
\begin{equation}
[q,p^{n-1}]¨=0=i(n-1)p^{n-2} \, \Rightarrow \,  p^{n-2}=0 \, , \ldots \, , p=0 \, \mbox{and} \, [q,p]=0 \neq i \, .
\end{equation}

\noindent The representation problem of infinitely many dimensions encountered above does not show up in the case of the
angular momentum algebra $\mathfrak{su}(2)$, where one has
\begin{equation}
[J_l , J_m]=i \varepsilon_{lmn} J_n \, , \quad \varepsilon_{lmn}=\frac{1}{2}(l-m)(m-n)(n-l) \, ,
\end{equation}
since these relations can be realized by the help of the
Pauli matrices $\{ \sigma_l \}_{l=1,2,3}$ by setting $J_l=\frac{1}{2} \sigma_l$
acting as linear operators on $\mathds{C}^2_\mathds{C}$.\\

\noindent Fortunately, there exists a so-called Weyl form of the CCR \cite{Weyl}, which uses unitary, i.e. bounded
and everywhere defined operators, only. Considering the unitary translation operator
$T_\beta$ acting on Lebesgue square integrable wave functions $\Psi \in \mathcal{L}^2(\mathds{R})$ according to
\begin{equation}
T_\beta \Psi(q) = \Psi(q-\beta) \overset{!}{=}
e^{-i \beta p} \Psi(q)  \, , \label{translation}
\end{equation}
the phase operator $e^{-i \alpha q}$ is also unitary and therefore
\begin{equation}
T_\beta e^{-i \alpha q} \Psi(q) = e^{-i \alpha (q -\beta)} \Psi(q-\beta) = e^{-i \alpha q}
e^{i \alpha \beta} T_\beta \Psi(q)
\end{equation}
or
\begin{equation}
T_\beta e^{-i \alpha q} = e^{ i \alpha \beta} \cdot  e^{-i \alpha q } T_\beta
\end{equation}
represents the Weyl form of the CCR for $F=1$, which is mathematically much more robust than
the better known form given by eqns. (\ref{ccrn}).
The self-adjoint momentum operator $p = -i \frac{d}{dq}$ is defined on a dense set $\mathcal{D}_{p}$
in the Hilbert space of wave functions $\mathcal{L}^2(\mathds{R})$,
where the expression
\begin{equation}
p = -i \lim_{\beta \rightarrow 0} \frac{id-T_\beta}{\beta} 
\end{equation}
makes sense, i.e.
\begin{equation}
\mathcal{D}_{p}=\{\Psi \,  \mbox{absolutely continuous}, \, d \Psi / dq \in \mathcal{L}^2(\mathds{R})\} \, ,
\end{equation}
and the originally formal exponential expression in eq. (\ref{translation}) becomes
well-defined on the whole Hilbert space $\mathcal{L}^2 ( \mathds{R})$.
The canonical commutation relation
\begin{equation}
[q, p] =i  \label{simplecomm}
\end{equation}
cannot hold on the whole Hilbert space $\mathcal{H}$, and eq. (\ref{simplecomm}) represents a dubious
statement as long the domain where it is defined is not discussed.

\end{document}